\documentclass[conference]{IEEEtran}
\IEEEoverridecommandlockouts
\usepackage{cite}
\usepackage{amsmath,amssymb,amsfonts}
\usepackage{algorithmic}
\usepackage{graphicx}
\usepackage{pgfplots}
\usepackage{booktabs}
\usepackage{tikz}
\usetikzlibrary{calc}
\usetikzlibrary{arrows.meta, positioning, shapes.multipart}
\pgfplotsset{compat=1.18}
\usepackage{hyperref}
\usepackage{textcomp}
\usetikzlibrary{positioning, fit, shapes.misc}
\usepackage{xcolor}
\def\BibTeX{{\rm B\kern-.05em{\sc i\kern-.025em b}\kern-.08em
    T\kern-.1667em\lower.7ex\hbox{E}\kern-.125emX}}
\begin{document}

\title{A Synthetic Conversational Smishing Dataset for Social Engineering Detection}

\author{\IEEEauthorblockN{Carl Lochstampfor}
\IEEEauthorblockA{\textit{Department of Cybersecurity} \\
\textit{Old Dominion University}\\
Norfolk, Virginia, United States\\
cloch001@odu.edu}
\and 
\IEEEauthorblockN{Ayan Roy}
\IEEEauthorblockA{\textit{School of Engineering and Computing} \\
\textit{Christopher Newport University}\\
Newport News, Virginia, United States \\
ayan.roy@cnu.edu}

}

\maketitle

\begin{abstract}

Smishing (SMS phishing) has emerged as a significant cybersecurity threat, particularly targeting elderly and cyber-unaware individuals, leading to substantial financial losses and erosion of trust. While prior research has proposed methods to classify individual text messages as smishing or benign, real-world attackers frequently employ multi-stage social engineering strategies, gradually grooming victims through extended conversational exchanges before extracting sensitive information. Despite the availability of datasets for single-message smishing detection, conversational smishing datasets remain largely absent, limiting research on multi-turn attack detection.
To address this gap, this paper introduces a synthetically generated dataset consisting of 3,201 multi-round labeled conversations that emulate realistic conversational smishing attacks. The dataset captures diverse attacker tactics and victim responses across multiple stages of engagement. To establish baseline performance, we evaluate eight models spanning traditional machine learning (Logistic Regression, Random Forest, Linear SVM, XGBoost) and transformer architectures (DistilBERT, Longformer), using both engineered conversational features and TF-IDF textual representations. Experimental results show that models leveraging TF-IDF features consistently outperform those using engineered features alone. The best-performing model, XGBoost with TF-IDF features, achieves an accuracy of 72.5\% and a macro F1 score of 0.691, outperforming both transformer models despite their contextual language understanding. Analysis reveals that input length limitations and limited training data size are the primary factors constraining transformer performance on this task. These results demonstrate the importance of lexical signals in conversational smishing detection and highlight the potential of the proposed dataset for advancing research in social engineering defense.

\end{abstract}

\begin{IEEEkeywords}
Social Engineering, elder fraud, synthetic data, scam detection, multi-agent LLM, conversation classification, Smishing Detection, Smishing Dataset
\end{IEEEkeywords}

\section{Introduction}

Short Message Service (SMS) phishing, commonly referred to as \textit{smishing}, has become one of the fastest growing forms of social engineering attacks. In these attacks, adversaries exploit the immediacy and perceived trustworthiness of text messaging to deceive victims into disclosing sensitive information or transferring funds, typically by impersonating trusted or reputable organizations~\cite{desolda2021human,xu2025malicious}. Recent cybersecurity reports indicate that smishing campaigns increasingly target elderly and technologically inexperienced individuals, who may be more susceptible to persuasion tactics and urgency-driven manipulation. The consequences of such attacks often include financial loss, identity theft, and erosion of trust in digital communication systems. Smishing attacks have increased by 18\% globally as of 2024, with 484,500 malicious smishing attempts reported in the United States, resulting in excess of \$400 million in financial losses ~\cite{keepnet2024smishing,cnet2024smishing}.

\begin{figure}[t]
\centering
\resizebox{\columnwidth}{!}{%
\begin{tikzpicture}[
    font=\scriptsize,
    phone/.style={
        draw=black!70,
        rounded corners=6pt,
        thick,
        minimum width=5.2cm,
        minimum height=9.3cm,
        fill=gray!6
    },
    screen/.style={
        draw=black!35,
        rounded corners=5pt,
        thick,
        minimum width=4.8cm,
        minimum height=8.0cm,
        fill=white
    },
    bubbleL/.style={
        draw=gray!45,
        rounded corners=4pt,
        fill=gray!15,
        align=left,
        text width=1.95cm,
        inner sep=3.5pt
    },
    bubbleR/.style={
        draw=blue!60!black,
        rounded corners=4pt,
        fill=blue!12,
        align=left,
        text width=1.75cm,
        inner sep=3.5pt
    }
]

\node[phone] (p1) at (0,0) {};
\node[screen] (s1) at (p1.center) {};

\draw[black!60, line width=0.7pt] ($(p1.north)+(-0.35,-0.22)$) -- ($(p1.north)+(0.35,-0.22)$);
\fill[black!60] ($(p1.north)+(0,-0.38)$) circle (0.03);

\node[font=\bfseries\scriptsize] at ($(p1.north)+(0,-0.5)$) {Single-Message Smishing};

\node[bubbleL, anchor=north west] (l1) at ($(s1.north west)+(0.18,-0.5)$)
{\textbf{Unknown Number}\\
Your bank account has been locked. Verify immediately at:\\
\texttt{secure-bank-}\\
\texttt{alert.com}};

\node[phone] (p2) at (5.8,0) {};
\node[screen] (s2) at (p2.center) {};

\draw[black!60, line width=0.7pt] ($(p2.north)+(-0.35,-0.22)$) -- ($(p2.north)+(0.35,-0.22)$);
\fill[black!60] ($(p2.north)+(0,-0.38)$) circle (0.03);

\node[font=\bfseries\scriptsize] at ($(p2.north)+(0,-0.5)$) {Conversational Grooming};

\node[bubbleL, anchor=north west] (r1) at ($(s2.north west)+(0.18,-0.45)$)
{\textbf{Unknown Number}\\
Hi Mrs. Carter, this is Anna from Medicare support.};

\node[bubbleR, anchor=north east] (r2) at ($(s2.north east)+(-0.18,-1.85)$)
{I did get something in the mail. What is this about?};

\node[bubbleL, anchor=north west] (r3) at ($(s2.north west)+(0.18,-3.15)$)
{There may be a verification issue with your coverage.};

\node[bubbleR, anchor=north east] (r4) at ($(s2.north east)+(-0.18,-4.55)$)
{Oh no, I don't want that. What do you need from me?};

\node[bubbleL, anchor=north west] (r5) at ($(s2.north west)+(0.18,-5.95)$)
{Please confirm your Medicare ID so I can verify your record.};

\end{tikzpicture}%
}
\caption{Comparison between single-message smishing and conversational grooming-based smishing. While the single-message attack (\textbf{on the left}) exhibits immediate malicious intent, conversational smishing (\textbf{on the right}) involves gradual trust-building and delayed elicitation of sensitive information across multiple turns, making detection significantly more challenging.}
\label{fig:single_vs_conversational_smishing}
\end{figure}
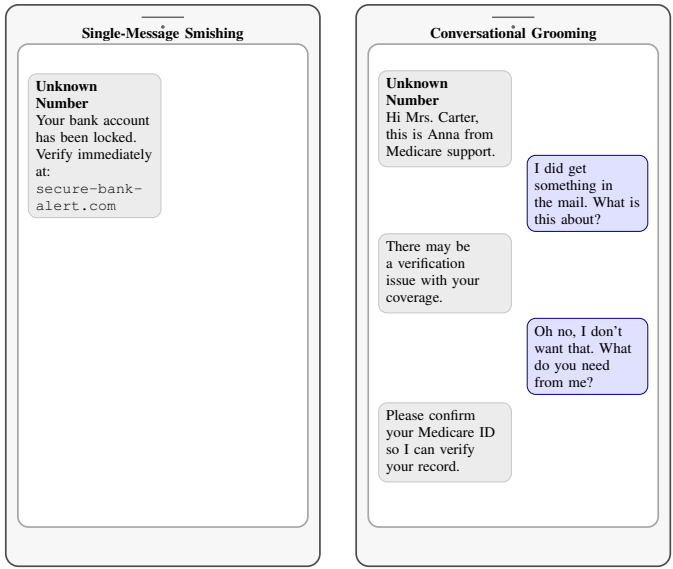

Most existing research on smishing detection \cite{alhuzali2026phishnet,birthriya2026mlsmishing,guangliang2026collaborative} focuses on classifying individual text messages as either malicious or benign. While this approach is useful for detecting isolated phishing attempts, it does not fully capture the behavior of modern attackers. In practice, many smishing attacks unfold as multi-step conversational interactions in which the attacker gradually builds trust with the victim before requesting sensitive information or payment~\cite{ent2025grooming}. A snapshot of such interactions is illustrated in Fig.~\ref{fig:single_vs_conversational_smishing}. These conversational attacks may involve multiple rounds of dialogue, emotional manipulation, and incremental extraction of personal or financial information. As a result, detection methods designed for single-message analysis may fail to capture the broader conversational context that reveals malicious intent.

Despite the growing prevalence of conversational social engineering attacks, publicly available datasets for studying multi-turn smishing interactions remain extremely limited. Most existing datasets consist of isolated SMS messages without conversational structure, which restricts the ability of researchers to study how attackers adapt their strategies over time or how victims respond during extended interactions. Furthermore, collecting real-world conversational smishing data presents significant privacy, ethical, and legal challenges, particularly when vulnerable populations such as elderly individuals are involved.

To address this gap, this paper introduces a synthetically generated conversational smishing dataset consisting of 3,201 multi-round labeled conversations across multiple scam categories. The dataset is designed to simulate realistic attacker–victim interactions and capture the dynamics of conversational social engineering attacks, including varied attacker strategies, victim responses, and outcomes. By leveraging a controlled multi-agent generation process, the dataset provides diverse conversational patterns while avoiding the ethical and privacy concerns associated with real-world data collection.
In addition to presenting the dataset, we establish baseline performance benchmarks using eight models spanning traditional machine learning (Logistic Regression, Random Forest, Linear Support Vector Machines, and XGBoost) and transformer architectures (DistilBERT and Longformer). We evaluate engineered conversational features, TF-IDF textual representations, and contextual embeddings to understand which approaches are most effective for detecting conversational smishing outcomes. Experimental results show that models leveraging TF-IDF features consistently outperform both engineered features and fine-tuned transformers. The best-performing configuration, XGBoost with TF-IDF features, achieves an accuracy of 72.5\% and a macro F1 score of 0.691 on the test set.

The main contributions of this paper are summarized as follows:

\begin{itemize}
\item We introduce a synthetically generated dataset of 3,201 multi-turn smishing conversations designed to model realistic conversational social engineering attacks.
\item We provide a structured labeling framework capturing different victim response outcomes within conversational attacks.
\item We evaluate eight baseline models --- six traditional ML classifiers and two transformer architectures --- using both engineered conversational features and TF-IDF textual representations.
\item We demonstrate that TF-IDF-based models outperform transformer models on this dataset, and analyze the contributing factors of input truncation and training data limitations.
\end{itemize}

The proposed dataset and baseline benchmarks establish a foundation for future research on conversational smishing detection and aim to support the development of more effective defensive strategies against social engineering attacks targeting vulnerable populations.

\section{Related Works}

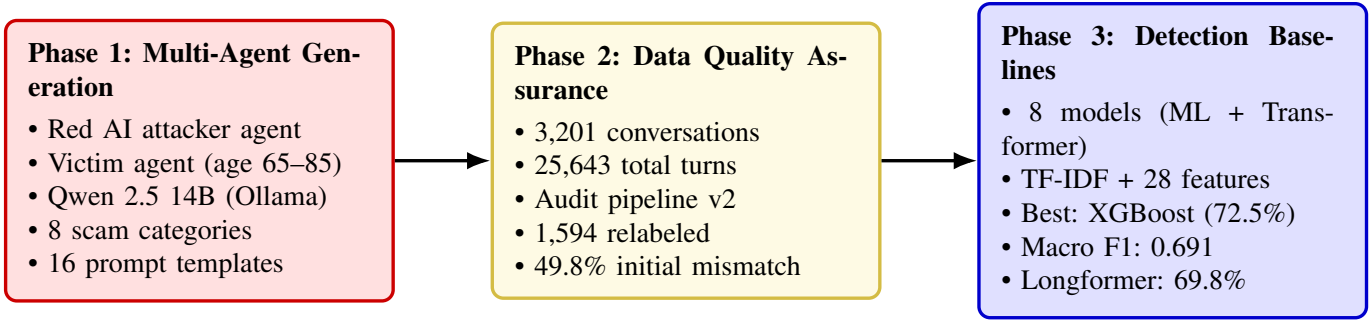
\begin{figure*}[!t]
\centering
\resizebox{\textwidth}{!}{%
\begin{tikzpicture}[
>=Latex,
node distance=1.2cm,
phase/.style={
rectangle,
rounded corners,
draw=#1!80!black,
fill=#1!12,
very thick,
minimum width=0.255\textwidth,
align=left,
inner sep=8pt
},
arrow/.style={->, very thick}
]

\node[phase=red] (p1) {
\begin{minipage}{0.23\textwidth}
\textbf{Phase 1: Multi-Agent Generation}

\smallskip
• Red AI attacker agent

• Victim agent (age 65--85)

• Qwen 2.5 14B (Ollama)

• 8 scam categories

• 16 prompt templates
\end{minipage}
};

\node[phase=yellow!80!orange, right=of p1] (p2) {
\begin{minipage}{0.23\textwidth}
\textbf{Phase 2: Data Quality Assurance}

\smallskip
• 3,201 conversations

• 25,643 total turns

• Audit pipeline v2

• 1,594 relabeled

• 49.8\% initial mismatch
\end{minipage}
};

\node[phase=blue, right=of p2] (p3) {
\begin{minipage}{0.23\textwidth}
\textbf{Phase 3: Detection Baselines}

\smallskip
• 8 models (ML + Transformer)

• TF-IDF + 28 features

• Best: XGBoost (72.5\%)

• Macro F1: 0.691

• Longformer: 69.8\%
\end{minipage}
};

\draw[arrow] (p1.east) -- (p2.west);
\draw[arrow] (p2.east) -- (p3.west);

\end{tikzpicture}%
}
\caption{COVA framework architecture showing the three-stage pipeline: multi-agent data generation, data quality auditing, and machine learning baseline evaluation.}
\label{fig:COVA_framework}
\end{figure*}

\subsection{Social Engineering Detection}

Detection of social engineering attacks has been studied across multiple modalities. In the SMS domain, Seo et al.~\cite{seo2024smishing} proposed on-device smishing classifiers resistant to text evasion, while Patra et al.~\cite{patra2023smsdect} developed prediction models combining machine learning with text analysis. However, these approaches address single-message classification and do not capture the multi-turn conversational dynamics characteristic of phone-based social engineering.

For telephone-based attacks, Derakhshan et al.~\cite{derakhshan2021detecting} introduced ASsET, a detection system based on \textit{scam signatures} --- sets of speech acts that characterize different scam types. Their approach uses word embeddings to identify whether the semantic content of a scam signature appears in a conversation. Wood et al.~\cite{wood2023scambaiting} analyzed scam-baiting calls from YouTube, developing a methodology to semi-automatically identify scam stages and scripts at scale, providing insight into the sequential structure of phone scams. Lansley et al.~\cite{lansley2020seader} developed SEADer++, a machine learning framework for detecting social engineering attacks in online environments, demonstrating that content-based features can effectively identify manipulative communication patterns.

Xu et al. \cite{xu2025malicious} proposed an ensemble learning based model by combining Random Forest and Support Vector Machine, which helps in classifying a message as malicious or legitimate.  

Jain et al. \cite{jain2025detecting} propose a smishing detection framework that integrates feature-based methods with advanced language models such as BERT. By evaluating both TF-IDF and contextual embeddings, their results demonstrate that BERT is more effective in capturing semantic and contextual relationships. Additionally, to improve classification performance, the original multi-class problem is reformulated into two binary classification tasks, each addressed using multi-layer neural networks.

Oswald et al.~\cite{oswald2022spotspam} introduced Spot Spam, an intention analysis-based approach for detecting SMS spam that leverages BERT embeddings. As a pre-trained language model, BERT provides high-quality contextual representations that capture semantic meaning effectively. The proposed method encodes SMS messages using BERT embeddings and identifies patterns indicative of spam. Additionally, Akande et al.~\cite{akande2022development} presented a smartphone-based system for real-time smishing detection using rule-based techniques.

\subsection{Synthetic Data and Multi-Agent Simulation}

Recent work has explored LLM-based approaches to simulate and counter social engineering. Basta et al.~\cite{basta2025botwars} presented ``Bot Wars,'' a framework using competing LLMs as scam-baiters against phone scams through simulated adversarial dialogues. Their two-layer prompt architecture enables demographically authentic victim personas, and they validated their synthetic dataset of 3,200 dialogues against 179 hours of human scam-baiting interactions. Kumarage et al.~\cite{kumarage2024sevism} proposed SE-VSim, an LLM-agentic framework for simulating social engineering attack mechanisms in multi-turn conversations, modeling victim agents with varying personality traits to assess susceptibility. Spokoyny et al.~\cite{spokoyny2025victim} developed CHATTERBOX, an LLM-based system that automates long-term engagement with online scammers, addressing pig-butchering and similar interactive scams.

\subsection{Gap in the Literature}



While prior work has advanced detection and simulation in related domains, several important gaps remain. First, publicly available multi-turn scam conversation datasets are scarce, with most comprehensive datasets maintained privately by telecom providers. Second, existing synthetic datasets primarily focus on single scam categories or text-based online scams, rather than telephone-based attacks spanning diverse categories. Third, there is a lack of datasets that explicitly model the conversational dynamics of elder-targeted scams with parameterized victim profiles. Herrera et al.~\cite{herrera2024bridging} highlight this protection gap, noting that older adults remain disproportionately vulnerable to AI-enhanced scams, while existing protective technologies have not kept pace.

To address these limitations, we introduce a publicly available, multi-category conversational smishing dataset focused on elder-targeted scenarios. The proposed dataset captures realistic multi-turn interactions and models attacker behavior across multiple conversational rounds aimed at financial exploitation. In addition, we provide baseline detection models to support future research on conversational smishing detection.

\section{Proposed Methodology}
\label{sec:methodology}

The \emph{COVA} (Cognitive Operations Virtual Assistant) framework comprises three phases: synthetic conversation generation, data quality assurance, and detection model training. Figure \ref{fig:COVA_framework} provides an overview of the complete pipeline. The framework is designed to be reproducible and extensible, with all code and generated data publicly available.

\subsection{Multi-Agent Conversation Generation}

Unlike single-prompt approaches that generate entire conversations from a monolithic template, our framework employs two independent LLM agents — an attacker and a victim — that exchange dialogue turns iteratively. This design produces more natural conversation dynamics, including realistic resistance patterns, escalation tactics, and varied outcomes.

\begin{itemize}
    \item Attacker Agent: Each attacker agent is configured with a scam type, a specific persona (name, claimed role, backstory), tactical parameters (urgency level, payment method, requested amount), and a knowledge level reflecting how much the attacker knows about the victim prior to the call. We model three knowledge tiers: no prior knowledge (50\%, cold calls), partial knowledge such as the victim’s name (30\%), and full knowledge including family details (20\%).
    
    \item Victim Agent: Victim agents are parameterized with demographic attributes (age 65–85, living situation), personality traits (trust level, scam awareness, tech savviness, emotional tendency), and a target outcome that guides but does not guarantee the conversation trajectory.

    \item Turn-Based Generation: The conversation proceeds as follows: (1) the victim initiates with a greeting, (2) the attacker delivers an opening line consistent with the scam type, (3) agents alternate turns with each response conditioned on the full conversation history, and (4) the conversation terminates when a natural endpoint is reached. Each turn is generated by a separate LLM inference call, preserving the independence of each agent’s perspective.

\end{itemize}

\subsection{Scam Category Coverage}
We selected eight scam categories that disproportionately affect elderly populations, based on the FBI’s Elder Fraud Report~\cite{fbi2023elder} and FTC complaint data. Table \ref{tab:scam_categories} summarizes the categories, key attacker tactics, and typical payment vectors. Each category has dedicated prompt templates for both attacker and victim agents, totaling 16 prompt templates.

\begin{table}[t]
\caption{Scam categories with attacker tactics and payment vectors}
\label{tab:scam_categories}
\centering
\footnotesize
\setlength{\tabcolsep}{3pt}
\begin{tabular}{p{0.24\columnwidth} p{0.36\columnwidth} p{0.20\columnwidth}}
\hline
\textbf{Category} & \textbf{Key Tactics} & \textbf{Payment} \\
\hline
Grandparent & Emotional manipulation, urgency & Wire, gift cards \\
Virtual Kidnapping & Terror, ransom demands & Wire transfer \\
Medicare Fraud & Authority, trust exploitation & Personal info \\
Romance & Emotional bonding, extraction & Wire, crypto \\
Government Impersonation & Fear of legal consequences & Gift cards, wire \\
Investment & Greed, FOMO, fake returns & Crypto, wire \\
Lottery & Excitement, fee extraction & Wire, prepaid \\
Bank Impersonation & Institutional trust, urgency & Credentials \\
\hline
\end{tabular}
\end{table}

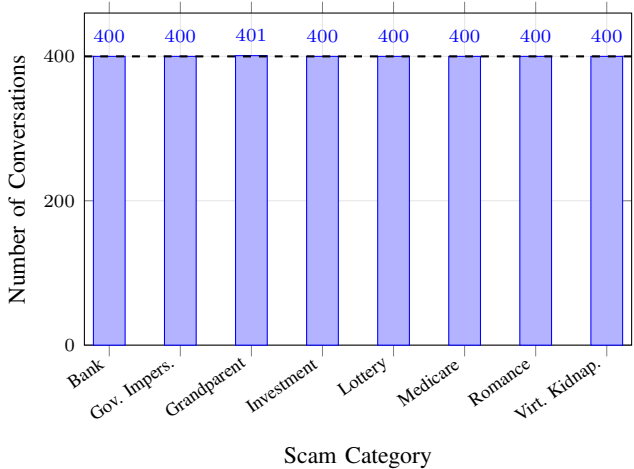
\begin{figure}[t]
\centering
\begin{tikzpicture}
\begin{axis}[
    ybar,
    bar width=12pt,
    width=\columnwidth,
    height=6cm,
    ymin=0,
    ymax=460,
    ylabel={Number of Conversations},
    xlabel={Scam Category},
    symbolic x coords={
        Bank,
        Gov.\ Impers.,
        Grandparent,
        Investment,
        Lottery,
        Medicare,
        Romance,
        Virt.\ Kidnap.
    },
    xtick=data,
    x tick label style={rotate=35, anchor=east, font=\scriptsize},
    yticklabel style={font=\scriptsize},
    xlabel style={font=\small},
    ylabel style={font=\small},
    nodes near coords,
    nodes near coords style={font=\scriptsize\bfseries, yshift=2pt},
    grid=major,
    grid style={gray!20},
    enlarge x limits=0.05,
    legend style={
        at={(0.98,0.98)},
        anchor=north east,
        font=\scriptsize,
        draw=gray!40,
        fill=white
    }
]

\addplot coordinates {
    (Bank,400)
    (Gov.\ Impers.,400)
    (Grandparent,401)
    (Investment,400)
    (Lottery,400)
    (Medicare,400)
    (Romance,400)
    (Virt.\ Kidnap.,400)
};

\draw[dashed, thick]
(axis description cs:0,{400/460}) --
(axis description cs:1,{400/460});
;

\end{axis}
\end{tikzpicture}
\caption{Distribution of synthetic conversations across eight scam categories as mentioned in Table \ref{tab:scam_categories} ($N=3{,}201$).}
\label{fig:dataset_distribution}
\end{figure}
\subsection{Local Inference Infrastructure}
We initially explored using commercial LLM APIs (Anthropic Claude) for conversation generation. However, the model’s safety guardrails prevented it from adequately roleplaying as a scam attacker, even when the research context was explicitly framed. We pivoted to local inference using Qwen 2.5 14B via Ollama, running on an NVIDIA RTX 4080 Super GPU. This approach offered three advantages: (1) no content restrictions for legitimate research, (2) zero API costs for bulk generation of over 3,000 conversations, and (3) complete data privacy with no data leaving the local machine.

\section{DATASET DESCRIPTION AND QUALITY ASSURANCE}
\subsection{Dataset Statistics}
The complete dataset comprises 3,201 synthetic multi-turn conversations distributed approximately equally across all eight scam categories (~400 per category). Each conversation is stored as a structured JSON file containing the full dialogue transcript, attacker and victim configurations, outcome labels, and generation metadata. The number of conversations corresponding to the category of scams as mentioned in Table \ref{tab:scam_categories} is highlighted in Figure \ref{fig:dataset_distribution}. 

\begin{figure}[]
\centering
\includegraphics[width=\columnwidth]{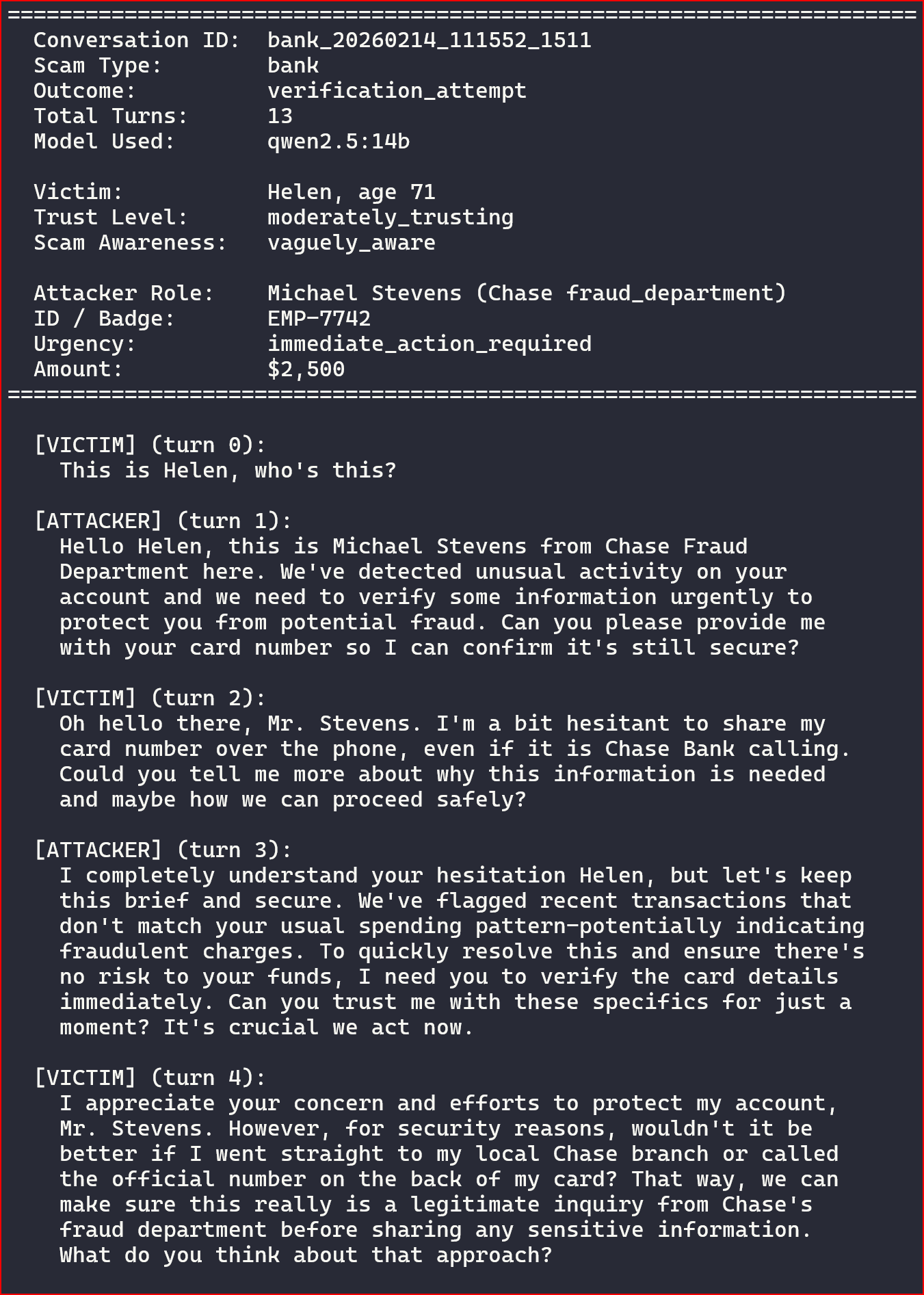}
\caption{Dataset Snapshot.}
\label{fig:conversation_snapshot}
\end{figure}

The dataset contains a total of 25,643 dialogue turns, with a mean of 16.2 turns per conversation and an average of 46.5 victim words per turn. All conversations are in English and simulate phone-based interactions. The summary of the dataset is provided in Table \ref{tab:dataset_stats}. The dataset and generation pipeline will be publicly released upon publication to support reproducibility and future research on conversational smishing detection. A snapshot of the dataset is shown in Figure \ref{fig:conversation_snapshot}. 

\begin{table}[t]
\centering
\caption{Summary statistics of the \emph{COVA} conversational smishing dataset}
\label{tab:dataset_stats}
\begin{tabular}{lc}
\toprule
\textbf{Metric} & \textbf{Value} \\
\midrule
Total conversations & 3,201 \\
Total conversation turns & 25,643 \\
Average turns per conversation & $\sim$16.2 \\
Number of scam categories & 8 \\
Victim age range & 65--85 \\
Prompt templates used & 16 \\
Relabeled conversations after audit & 1,594 \\
Initial label mismatch rate & 49.8\% \\
\bottomrule
\end{tabular}
\end{table}

\subsection{Outcome Classification}
The initial generation pipeline assigned one of five fine-grained outcome labels: successful\_scam, partial\_compliance, verification\_attempt, scam\_detected, and quick\_rejection. Because the scam\_detected ($n=67$) and quick\_rejection ($n=16$) classes contain too few samples for reliable classification, we collapse the five labels into three classes for modeling:

\begin{itemize}
    \item Complied: The victim fully complies with the attacker’s requests. Corresponds to the original successful\_scam label (n=534).
    \item Partial: The victim engages but does not fully commit. Corresponds to the original partial\_compliance label (n=1,748).
    \item Rejected: The victim rejects the scammer or terminates the call. Merges verification\_attempt (n=836), scam\_detected (n=67), and quick\_rejection (n=16) (combined n=919).
    
\end{itemize}
Average conversation length varies by outcome: quick rejections average 4.5 turns, while successful scams average 17.0 turns, reflecting the extended engagement required for full compliance.

\subsection{Label Quality Audit}
Initial outcome labels were assigned by a rule-based classifier during generation. Upon systematic validation, we discovered a 49.8\% mismatch rate between the assigned labels and the actual conversation content. Of the 3,201 conversations audited, only 1,290 (40.3\%) were confirmed correct; 1,594 (49.8\%) required relabeling; and 317 (9.9\%) were skipped due to low confidence. The primary failure modes were:

\begin{itemize}
    \item False compliance detection: Phrases like “I don’t have my credit card right now” were incorrectly classified as compliance, when the context indicated refusal.
    \item Missed successful scams: Victims who explicitly committed to action were labeled as partial\_compliance because the classifier failed to recognize strong commitment language.
    \item Context-insensitive keyword matching: The word “scam” in cautious victim phrases triggered a scam\_detected label even when the victim was still engaging.
\end{itemize}

\textbf{Correction Pipeline: } We developed an improved audit pipeline (v2) incorporating negation-aware matching, scam-type-specific heuristics, and weighted analysis of the final conversation turns. Of the 1,594 relabeled conversations, 1,339 (84.0\%) were medium-confidence corrections and 255 (16.0\%) were high-confidence corrections. Relabeling affected all eight scam categories, with grandparent scams (232 changes), romance scams (225), and virtual kidnapping (218) requiring the most corrections. Original labels were preserved in metadata for traceability. Manual validation of 28 stratified samples confirmed corrected labels aligned with human judgment in over 90\% of cases.



\section{Experimentation}

\subsection{Experimental Setup}

\subsubsection{Data Splits} 
The corrected dataset was partitioned into stratified train (80\%), validation (10\%), and test (10\%) splits using a random seed of 42, maintaining proportional representation of both scam categories and outcome classes. This yielded 2,556 training conversations (20,490 turns), 321 validation conversations (2,561 turns), and 324 test conversations (2,592 turns).

\subsubsection{Feature Representations}
We compared two feature extraction approaches: (1) \textbf{Engineered Features} — 28 numeric features including conversation length metrics, behavioral indicators (resistance/compliance counts and ratios), temporal patterns, financial signals, victim demographics, and stylistic features; and (2) \textbf{TF-IDF Features} — Term Frequency–Inverse Document Frequency vectors computed over the full conversation text using unigrams and bigrams with a maximum of 5,000 features and sublinear TF scaling, combined with the 28 engineered features.

\subsubsection{Models}
We trained six scikit-learn classifiers and fine-tuned two transformer models, spanning diverse algorithmic families, as summarized in Table \ref{tab:models_features}.

\begin{table}[t]
\caption{Models evaluated with feature types}
\label{tab:models_features}
\centering
\footnotesize
\setlength{\tabcolsep}{3pt}
\begin{tabular}{p{0.24\columnwidth} p{0.27\columnwidth} p{0.28\columnwidth}}
\hline
\textbf{Model} & \textbf{Feature Type} & \textbf{Description} \\
\hline
Logistic Regression & TF-IDF + features & Linear, L2 regularization \\
Random Forest & Features only & 100-tree ensemble \\
Random Forest & TF-IDF + features & 100-tree ensemble, combined \\
Linear SVM & TF-IDF + features & Linear kernel \\
XGBoost & Features only & Gradient boosting \\
XGBoost & TF-IDF + features & Gradient boosting, combined \\
DistilBERT & Contextual embeddings & Transformer, 66M params \\
Longformer & Contextual embeddings & Transformer, 148M params \\
\hline
\end{tabular}
\end{table}

\subsection{Baseline Results}

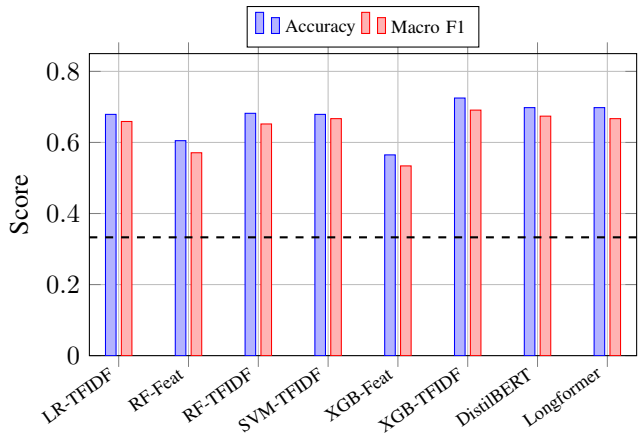
\begin{figure}[t]
\centering
\begin{tikzpicture}
\begin{axis}[
ybar,
bar width=4pt,
width=\columnwidth,
height=5.6cm,
ymin=0,
ymax=0.85,
ylabel={Score},
symbolic x coords={
LR-TFIDF,
RF-Feat,
RF-TFIDF,
SVM-TFIDF,
XGB-Feat,
XGB-TFIDF,
DistilBERT,
Longformer
},
xtick=data,
x tick label style={rotate=35,anchor=east,font=\scriptsize},
legend style={
at={(0.5,1.02)},
anchor=south,
legend columns=2,
font=\scriptsize
},
grid=major,
enlarge x limits=0.06
]

\addplot coordinates {
(LR-TFIDF,0.679)
(RF-Feat,0.605)
(RF-TFIDF,0.682)
(SVM-TFIDF,0.679)
(XGB-Feat,0.565)
(XGB-TFIDF,0.725)
(DistilBERT,0.698)
(Longformer,0.698)
};
\addlegendentry{Accuracy}

\addplot coordinates {
(LR-TFIDF,0.659)
(RF-Feat,0.571)
(RF-TFIDF,0.652)
(SVM-TFIDF,0.667)
(XGB-Feat,0.534)
(XGB-TFIDF,0.691)
(DistilBERT,0.674)
(Longformer,0.667)
};
\addlegendentry{Macro F1}

\draw[dashed, thick]
(axis description cs:0,{0.333/0.85}) --
(axis description cs:1,{0.333/0.85});

\end{axis}
\end{tikzpicture}

\caption{Model comparison showing accuracy and macro F1 for all eight models on 3-class outcome prediction. The dashed horizontal line represents the random baseline (0.333).}
\label{fig:baseline_results}
\end{figure}

Fig.~\ref{fig:baseline_results} presents the performance of all eight models on the held-out test set. The best-performing model was XGBoost with TF-IDF + engineered features, achieving 72.5\% accuracy and a macro F1 score of 0.691. This substantially outperformed all other configurations, including both transformer models.

The performance gap between XGBoost configurations is striking: XGBoost with TF-IDF achieved 72.5\% accuracy compared to only 56.5\% with engineered features alone --- an improvement of 16.0 percentage points. This demonstrates that lexical patterns carry critical discriminative signal that hand-crafted features fail to capture.

\subsection{Error Analysis}

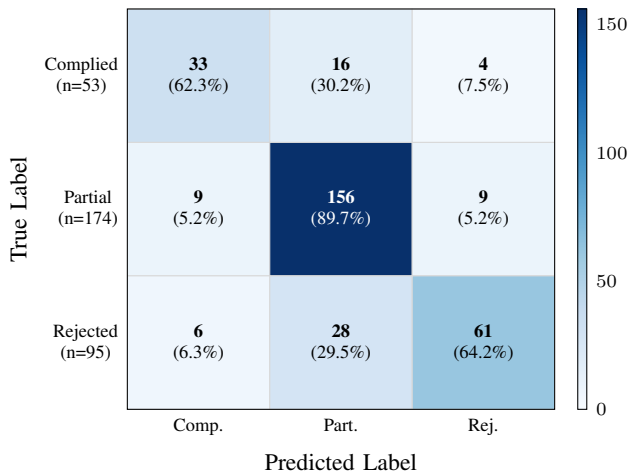
\begin{figure}[t]
\centering
\begin{tikzpicture}
\begin{axis}[
    width=0.82\columnwidth,
    height=0.78\columnwidth,
    view={0}{90},
    xmin=-0.5, xmax=2.5,
    ymin=-0.5, ymax=2.5,
    axis on top,
    enlargelimits=false,
    xtick={0,1,2},
    ytick={0,1,2},
    xticklabels={{Comp.},{Part.},{Rej.}},
    yticklabels={{Rejected\\(n=95)},{Partial\\(n=174)},{Complied\\(n=53)}},
    xlabel={Predicted Label},
    ylabel={True Label},
    tick style={draw=none},
    xlabel style={font=\small},
    ylabel style={font=\small},
    xticklabel style={font=\scriptsize},
    yticklabel style={font=\scriptsize, align=center},
    point meta min=0,
    point meta max=156,
    colorbar,
    colorbar style={
        width=0.12cm,
        yticklabel style={font=\scriptsize},
        tick style={draw=none}
    },
    colormap={customblue}{
        rgb255(0cm)=(247,251,255)
        rgb255(1cm)=(222,235,247)
        rgb255(2cm)=(198,219,239)
        rgb255(3cm)=(158,202,225)
        rgb255(4cm)=(107,174,214)
        rgb255(5cm)=(66,146,198)
        rgb255(6cm)=(33,113,181)
        rgb255(7cm)=(8,81,156)
        rgb255(8cm)=(8,48,107)
    }
]

\addplot [
    matrix plot*,
    mesh/rows=3,
    point meta=explicit,
    draw=gray!30
] table [meta=value] {
x y value
0 2 33
1 2 16
2 2 4
0 1 9
1 1 156
2 1 9
0 0 6
1 0 28
2 0 61
};

\node[font=\scriptsize, align=center] at (axis cs:0,2) {\textbf{33}\\(62.3\%)};
\node[font=\scriptsize, align=center] at (axis cs:1,2) {\textbf{16}\\(30.2\%)};
\node[font=\scriptsize, align=center] at (axis cs:2,2) {\textbf{4}\\(7.5\%)};

\node[font=\scriptsize, align=center] at (axis cs:0,1) {\textbf{9}\\(5.2\%)};
\node[font=\scriptsize, align=center, text=white] at (axis cs:1,1) {\textbf{156}\\(89.7\%)};
\node[font=\scriptsize, align=center] at (axis cs:2,1) {\textbf{9}\\(5.2\%)};

\node[font=\scriptsize, align=center] at (axis cs:0,0) {\textbf{6}\\(6.3\%)};
\node[font=\scriptsize, align=center] at (axis cs:1,0) {\textbf{28}\\(29.5\%)};
\node[font=\scriptsize, align=center] at (axis cs:2,0) {\textbf{61}\\(64.2\%)};

\end{axis}
\end{tikzpicture}
\caption{Confusion matrix for XGBoost + TF-IDF on the 3-class test set ($n=324$). Percentages indicate row-wise recall.}
\label{fig:confusion_matrix}
\end{figure}

The confusion matrix depicted in Figure \ref{fig:confusion_matrix} for the \textbf{XGBoost + TF-IDF } reveals the following patterns:

The \textit{partial} class achieves the highest recall (89.7\%), benefiting from larger representation and uniform language patterns. The \textit{complied} class achieves 62.3\% recall, with most errors involving misclassification as \textit{partial} (16 of 53 cases) --- expected since conversations where victims eventually comply contain extended engagement similar to partial compliance. The \textit{rejected} class achieves 64.2\% recall, with 28 of 95 cases misclassified as \textit{partial}, reflecting lexical overlap between extensive engagement before rejection and engagement without commitment. The model achieves its highest precision on \textit{rejected} (82.4\%), indicating high confidence when predicting rejection.

\subsection{Five-Class Analysis}

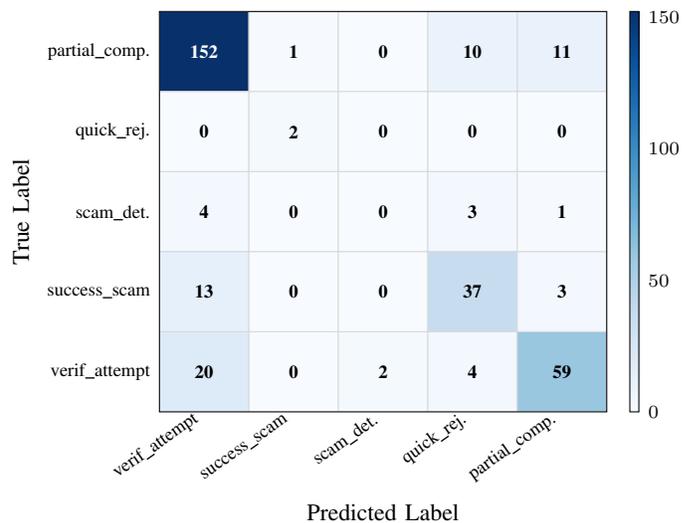
\begin{figure}[t]
\centering
\begin{tikzpicture}
\begin{axis}[
    width=0.85\columnwidth,
    height=0.78\columnwidth,
    view={0}{90},
    xmin=-0.5, xmax=4.5,
    ymin=-0.5, ymax=4.5,
    axis on top,
    enlargelimits=false,
    xtick={0,1,2,3,4},
    ytick={0,1,2,3,4},
    xticklabels={{verif\_attempt},{success\_scam},{scam\_det.},{quick\_rej.},{partial\_comp.}},
    yticklabels={{verif\_attempt},{success\_scam},{scam\_det.},{quick\_rej.},{partial\_comp.}},
    xlabel={Predicted Label},
    ylabel={True Label},
    tick style={draw=none},
    xlabel style={font=\small},
    ylabel style={font=\small},
    xticklabel style={font=\scriptsize, rotate=35, anchor=east},
    yticklabel style={font=\scriptsize},
    point meta min=0,
    point meta max=152,
    colorbar,
    colorbar style={
        width=0.12cm,
        yticklabel style={font=\scriptsize},
        tick style={draw=none}
    },
    colormap={customblue}{
        rgb255(0cm)=(247,251,255)
        rgb255(1cm)=(222,235,247)
        rgb255(2cm)=(198,219,239)
        rgb255(3cm)=(158,202,225)
        rgb255(4cm)=(107,174,214)
        rgb255(5cm)=(66,146,198)
        rgb255(6cm)=(33,113,181)
        rgb255(7cm)=(8,81,156)
        rgb255(8cm)=(8,48,107)
    }
]

\addplot [
    matrix plot*,
    mesh/rows=5,
    point meta=explicit,
    draw=gray!30
] table [meta=value] {
x y value
0 4 152
1 4 1
2 4 0
3 4 10
4 4 11

0 3 0
1 3 2
2 3 0
3 3 0
4 3 0

0 2 4
1 2 0
2 2 0
3 2 3
4 2 1

0 1 13
1 1 0
2 1 0
3 1 37
4 1 3

0 0 20
1 0 0
2 0 2
3 0 4
4 0 59
};

\node[font=\scriptsize, text=white] at (axis cs:0,4) {\textbf{152}};
\node[font=\scriptsize] at (axis cs:1,4) {\textbf{1}};
\node[font=\scriptsize] at (axis cs:2,4) {\textbf{0}};
\node[font=\scriptsize] at (axis cs:3,4) {\textbf{10}};
\node[font=\scriptsize] at (axis cs:4,4) {\textbf{11}};

\node[font=\scriptsize] at (axis cs:0,3) {\textbf{0}};
\node[font=\scriptsize] at (axis cs:1,3) {\textbf{2}};
\node[font=\scriptsize] at (axis cs:2,3) {\textbf{0}};
\node[font=\scriptsize] at (axis cs:3,3) {\textbf{0}};
\node[font=\scriptsize] at (axis cs:4,3) {\textbf{0}};

\node[font=\scriptsize] at (axis cs:0,2) {\textbf{4}};
\node[font=\scriptsize] at (axis cs:1,2) {\textbf{0}};
\node[font=\scriptsize] at (axis cs:2,2) {\textbf{0}};
\node[font=\scriptsize] at (axis cs:3,2) {\textbf{3}};
\node[font=\scriptsize] at (axis cs:4,2) {\textbf{1}};

\node[font=\scriptsize] at (axis cs:0,1) {\textbf{13}};
\node[font=\scriptsize] at (axis cs:1,1) {\textbf{0}};
\node[font=\scriptsize] at (axis cs:2,1) {\textbf{0}};
\node[font=\scriptsize] at (axis cs:3,1) {\textbf{37}};
\node[font=\scriptsize] at (axis cs:4,1) {\textbf{3}};

\node[font=\scriptsize] at (axis cs:0,0) {\textbf{20}};
\node[font=\scriptsize] at (axis cs:1,0) {\textbf{0}};
\node[font=\scriptsize] at (axis cs:2,0) {\textbf{2}};
\node[font=\scriptsize] at (axis cs:3,0) {\textbf{4}};
\node[font=\scriptsize] at (axis cs:4,0) {\textbf{59}};

\end{axis}
\end{tikzpicture}
\caption{Five-class confusion matrix for XGBoost + TF-IDF (test set). Minority classes \textit{scam\_detected} and \textit{quick\_rejection} show near-zero recall.}
\label{fig:five_class_confusion}
\end{figure}

We also evaluated all models on the original five-class labels. The best model remained XGBoost + TF-IDF, achieving 72.5\% accuracy but a substantially lower macro F1 of 0.614. The drop is attributable to near-zero performance on minority classes as depicted in Figure \ref{fig:five_class_confusion}: \textit{quick\_rejection} ($n=2$ in test) and \textit{scam\_detected} ($n=8$ in test) have insufficient samples for reliable classification. This motivated the 3-class collapsed evaluation as primary.

\subsection{Feature Importance}

The top features identified by XGBoost + TF-IDF include both engineered and lexical features. The top-ranked engineered features were \textit{victim\_turn\_count} and \textit{attacker\_turn\_count}, confirming that conversation length is a strong predictor. Among TF-IDF features, terms related to verification behavior (e.g., ``official,'' ``hang up,'' ``the card'') and scam awareness (e.g., ``be aware,'' ``trick,'' ``due diligence'') were most discriminative.

\subsection{Transformer Results}

To investigate whether contextual language models can improve upon bag-of-words representations, we fine-tuned two transformer architectures on the same 3-class task.

\subsubsection{DistilBERT} We fine-tuned DistilBERT-base-uncased (66M parameters) for 8 epochs with a learning rate of $2 \times 10^{-5}$, batch size 16, and balanced class weights. DistilBERT has a maximum input length of 512 tokens, but our conversations average 377 tokens with a maximum of 874. We evaluated three truncation strategies: head-only (default, keeping the first 512 tokens), tail-only (keeping the last 512 tokens). The tail-only (v2) strategy performed best, achieving 69.8\% accuracy and 0.674 macro F1 on the test set. The tail-only strategy retains the outcome-critical final turns, and consistently outperformed head-only configurations across validation runs.

\subsubsection{Longformer} To isolate the impact of truncation, we fine-tuned Longformer-base (148M parameters, 4,096-token capacity) with a maximum input length of 1,024 tokens --- sufficient to fit all conversations without any truncation. We used a batch size of 4, learning rate of $2 \times 10^{-5}$, and trained for 8 epochs with balanced class weights. Longformer achieved 69.8\% accuracy and 0.667 macro F1, substantially outperforming all DistilBERT configurations but falling short of XGBoost + TF-IDF.

Table~\ref{tab:full_results} summarizes the complete results across all eight model configurations. XGBoost + TF-IDF remains the best-performing model, with Longformer as the strongest transformer baseline.

\begin{table}[t]
\caption{Complete 3-class test results ($n=324$). Best in bold.}
\centering
\footnotesize
\setlength{\tabcolsep}{4pt}
\begin{tabular}{lccc}
\toprule
\textbf{Model} & \textbf{Accuracy} & \textbf{F1 Macro} & \textbf{F1 Weighted} \\
\midrule
\textbf{XGBoost + TF-IDF} & \textbf{0.725} & \textbf{0.691} & \textbf{0.718} \\
Longformer (1024 tok.) & 0.698 & 0.667 & 0.701 \\
DistilBERT (tail, v2) & 0.698 & 0.674 & 0.695 \\
Linear SVM + TF-IDF & 0.679 & 0.667 & 0.678 \\
Logistic Reg. + TF-IDF & 0.679 & 0.659 & 0.676 \\
Random Forest + TF-IDF & 0.682 & 0.652 & 0.678 \\
Random Forest (feat.) & 0.605 & 0.571 & 0.593 \\
XGBoost (feat. only) & 0.565 & 0.534 & 0.557 \\
\bottomrule
\end{tabular}
\label{tab:full_results}
\end{table}
\section{Discussion}

\subsection{Dataset Contribution}

The primary contribution of this work is the synthetic dataset itself. With 3,201 conversations across eight scam categories, this represents --- to our knowledge --- the largest publicly available multi-turn social engineering conversation dataset targeting elderly populations. The multi-agent generation approach produces conversations with natural turn-taking dynamics, varied resistance patterns, and realistic outcomes that would be impractical to collect from real-world interactions due to privacy and ethical constraints.

The 49.8\% initial label mismatch rate underscores an important finding: automated outcome labeling of conversational data is substantially harder than labeling single messages. Context-dependent phrases, negation, and the temporal progression of a conversation all contribute to labeling difficulty. Our improved audit pipeline demonstrates that systematic, heuristic-based correction can substantially improve label quality, but also highlights the need for future work on more sophisticated labeling approaches.

\subsection{Baseline Analysis}

The 72.5\% accuracy achieved by XGBoost + TF-IDF on a 3-class prediction task represents a meaningful baseline above the 33.3\% random chance level, but leaves substantial room for improvement. The dominance of TF-IDF over engineered features (16.0-point accuracy gap for XGBoost) demonstrates that lexical patterns carry critical discriminative signal that hand-crafted features fail to capture.

Contrary to initial expectations, transformer-based models did not outperform TF-IDF-based classifiers on this dataset. We attribute this to two factors. First, \textit{input length limitations}: DistilBERT's 512-token limit forces truncation of conversations that average 377 tokens with a maximum of 874. This truncation discards critical late-conversation signals --- the compliance or rejection indicators that determine the outcome label. The progressive improvement from DistilBERT head-only to tail-only (69.8\%) to Longformer with no truncation (69.8\%) confirms that context window size directly impacts performance. Second, \textit{dataset scale}: with only 2,556 training conversations, transformer models lack sufficient data for effective fine-tuning. Longformer's training loss fell to 0.18 by epoch 8 while validation loss climbed to 2.30, exhibiting clear overfitting characteristic of data-limited regimes. In contrast, XGBoost operates over pre-computed TF-IDF features and does not need to learn language representations from scratch, making it inherently more sample-efficient.

These findings suggest that for small conversational datasets, traditional ML with TF-IDF features provides a strong and practical baseline. Transformer models would likely benefit from larger training corpora ($>$5,000 conversations) and longer-context architectures, which we identify as directions for future work.

The confusion between \textit{partial} and \textit{rejected} classes remains a challenge across all model families. Victims who engage at length before ultimately rejecting a scam produce dialogue that is lexically similar to those who engage without ever fully committing. This boundary may require turn-level temporal modeling to resolve effectively.

\subsection{Limitations}

Several limitations should be acknowledged. First, the conversations are synthetic and may not fully capture the linguistic patterns of real scam interactions, particularly the emotional dynamics of actual elderly victims. Second, the dataset is English-only and US-centric. Third, the label correction pipeline was validated on a relatively small stratified sample (28 conversations) and may contain residual errors --- 317 low-confidence conversations (9.9\%) were left unchanged. Fourth, the current baseline models treat each conversation as a complete document, whereas a real-time detection system would need to classify incrementally. Fifth, the 3-class collapse merges meaningful behavioral distinctions that may be important for intervention systems. Sixth, the dataset size of 3,201 conversations may be insufficient for effective transformer fine-tuning, as evidenced by the overfitting observed in both DistilBERT and Longformer experiments.

\section{Conclusion and Future Work}
This paper presents three contributions to the field of AI-driven social engineering defense. First, we introduce a multi-agent LLM framework for generating synthetic scam conversations and the resulting dataset of 3,201 conversations across eight elder-targeted scam categories --- the first publicly available resource of its kind. Second, we establish detection baselines using eight models spanning traditional ML and transformer architectures, with XGBoost + TF-IDF achieving 72.5\% accuracy and 0.691 macro F1 on 3-class outcome prediction. Third, we provide an empirical analysis showing that TF-IDF-based models outperform fine-tuned transformers on conversational datasets of this scale, identifying input length limitations and training data size as the primary constraining factors.

Future work will proceed along three directions. First, we plan to expand the dataset to over 10,000 conversations to investigate whether larger training corpora enable transformer models to surpass traditional ML baselines. Second, we will explore incremental classification --- predicting conversation outcomes from partial transcripts to enable real-time intervention. Third, we intend to expand the dataset with additional scam categories and non-English conversations to improve generalizability. The COVA conversational smishing dataset, generation pipeline, and detection baselines will be made publicly available upon acceptance of this paper.

\bibliographystyle{IEEEtran}
\bibliography{reference}
\end{document}